\documentclass[twocolumn,showpacs,amsmath,amssymb,pre]{revtex4}

\usepackage{graphics}
\usepackage{epsfig}

\begin{document}

\title{Kosmotropes and chaotropes: modelling preferential exclusion, 
binding and aggregate stability}

\author{Susanne Moelbert}
\affiliation{Institut de th\'eorie des ph\'enom\`enes physiques, Ecole 
polytechnique f\'ed\'erale de Lausanne, CH-1015 Lausanne, Switzerland}

\author{B. Normand} 
\affiliation{D\'epartement de Physique, Universit\'e de Fribourg,  
CH-1700 Fribourg, Switzerland} 

\author{Paolo De Los Rios}
\affiliation{Institut de th\'eorie des ph\'enom\`enes physiques, Ecole 
polytechnique f\'ed\'erale de Lausanne, CH-1015 Lausanne, Switzerland}
\affiliation{INFM UdR - Politecnico, Corso Duca degli Abruzzi 24, 10129 
Torino, Italy.} 

\date{\today}

\begin{abstract}

Kosmotropic cosolvents added to an aqueous solution promote the 
aggregation of hydrophobic solute particles, while chaotropic cosolvents 
act to destabilise such aggregates. We discuss the mechanism for these 
phenomena within an adapted version of the two-state Muller-Lee-Graziano 
model for water, which provides a complete description of the ternary 
water/cosolvent/solute system for small solute particles. This model 
contains the dominant effect of a kosmotropic substance, which is to 
enhance the formation of water structure. The consequent preferential 
exclusion both of cosolvent molecules from the solvation shell of 
hydrophobic particles and of these particles from the solution leads 
to a stabilisation of aggregates. By contrast, chaotropic substances 
disrupt the formation of water structure, are themselves preferentially 
excluded from the solution, and thereby contribute to solvation of 
hydrophobic particles. We use Monte Carlo simulations to demonstrate 
at the molecular level the preferential exclusion or binding of cosolvent 
molecules in the solvation shell of hydrophobic particles, and the 
consequent enhancement or suppression of aggregate formation. We 
illustrate the influence of structure-changing cosolvents on effective 
hydrophobic interactions by modelling qualitatively the kosmotropic 
effect of sodium chloride and the chaotropic effect of urea. 

\end{abstract}


\maketitle

\section{Introduction}

Most processes in living organisms are adjusted to function in a rather 
limited range of physiological conditions, and important deviations, 
such as high concentrations of dangerous substances, are generally 
expected to preclude life. Nevertheless, many living systems survive 
stresses of this kind and exist in hostile environments. One common 
adaptation strategy is to modify the properties of the solvent, which 
is usually water, in such a way as to exclude the undesirable solutes 
from solution~\cite{galinski2}. Water is modified by relatively high 
concentrations of stabilising solutes (cosolvents), which remain 
compatible with the metabolism of the cell even at very high 
concentrations (therefore they are also referred to as `compatible 
osmolytes')~\cite{yancey,somero}. These cosolvents neutralise dangerous 
solutes by decreasing their solubility and enhancing the formation of 
their aggregates. Such cosolvents are known as promoters of the water 
structure and are therefore referred to as kosmotropes (`kosmo-trope' 
$=$ order maker). Their stabilising function for proteins and aggregates 
of hydrophobic solute particles, and their importance for the osmotic 
balance in cells, has generated a growing interest in the physical origin 
of the kosmotropic effect. Many recent investigations have focused on 
the ability of kosmotropic cosolvents to influence solute solubility in 
aqueous media~\cite{wiggins,galinski,arakawa,timasheff3,rszjk,rbb}. 

Of equal interest and importance is the ability of chaotropic cosolvents 
to increase the solubility of non-polar solute particles in aqueous 
solutions~\cite{schellman, timasheff2, tovchi, oro, feng, plumridge}. 
For certain systems the solubility may be enhanced by several orders 
of magnitude~\cite{nozaki1, nozaki1a, nandi}, leading to a complete 
destabilisation of solute aggregates and, in the case of protein 
solutions, to a complete denaturation and loss of function. A 
significant number of substances display this property, notably 
(under most conditions and for the majority of solute species) urea, 
which is frequently used in highly concentrated ({\it c}.~0.5-10 M) 
solutions as a protein denaturant. Chaotropic cosolvents (`chao-tropic' 
$=$ order-breaking) are less polar than water and thus break hydrogen 
bonds between water molecules, suppressing water structure 
formation~\cite{galinski,walrafen,wetlaufer,roseman}. However, the 
exact physical mechanism for the changes in water structure, and for 
the consequent destabilising action of chaotropic cosolvents, is at 
present not fully understood. 
 
\begin{figure}[t!]
\includegraphics[width=8.5cm]{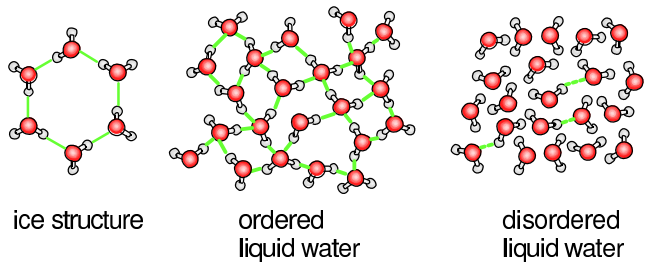}
\caption{Schematic illustration of water structure as a consequence of 
the formation of more or less extended networks of hydrogen bonds.}
\label{Figws}
\end{figure}

Because the origin of the kosmotropic and chaotropic effects appears to 
lie primarily in their influence on the solvent, rather than in direct 
interactions between cosolvent and solute~\cite{kita,franks,creighton}, 
a microscopic description must begin with the unique properties of the 
aqueous medium. Water molecules have the ability to form strong, 
intermolecular hydrogen bonds, and pure, liquid water may form extended 
hydrogen-bonded networks, becoming highly ordered (Fig.~\ref{Figws}). 
Although the insertion of a hydrophobic molecule leads to a destruction 
of local hydrogen-bond structure, at low temperatures water molecules 
are found to rearrange in a cage-like configuration around small 
solute particles and around high-curvature regions of larger ones. 
The orientation imposed on the water molecules at the interface 
with the particle results in the cage hydrogen bonds being slightly 
stronger than before, and causing a net reduction of energy during 
the insertion process~\cite{frank,yaacobi,privalov,deJong,pertsemil}. 
However, at higher temperatures the free energy of the system is 
decreased by reducing the local restructuring of water, thus increasing 
the entropy of the solvent molecules, which drives the aggregation of 
hydrophobic particles through a minimisation of their total surface 
exposed to water. This competition between enthalpic and entropic 
effects in the solvent is fundamental to the phenomenology of 
liquid-liquid demixing processes. The effective hydrophobic interaction 
between small, non-polar solute particles is thus thought to be primarily 
solvent-induced, {\it i.e.}~to be a consequence of changes in the ordering 
of water molecules rather than being controlled by direct water-solute 
interactions~\cite{kauzmann,tanford,stillinger80,ben-naim2,ludwig}. In 
this description, the primary contribution to the action of 
structure-changing cosolvents, which are generally highly soluble and 
uncharged in physiological conditions, then depends on their ability to 
alter this local ordering of liquid water. 

Kosmotropic cosolvents, such as sucrose and betaine, are more 
polar than water and act to enhance its structure due to their ability 
to form hydrogen bonds~\cite{galinski}. For the same reason they interact 
with water molecules rather than with non-polar solute particles, which 
leads to an effective preferential exclusion from the solvation shell of 
hydrophobic molecules~\cite{galinski2,arakawa,creighton}, and thus to a 
stronger net repulsion between solute and solvent. In the presence of 
kosmotropic cosolvents, structural arrangement of the water-cosolvent 
mixture is enthalpically favourable compared to a cage-like organisation 
around hydrophobic solute particles. Solute molecules are thus pushed 
together to minimise their total exposed surface, which results in an 
enhancement of hydrophobic aggregation. The same process leads to a 
stabilisation of native protein configurations, in spite of the fact that 
kosmotropic substances have no net charge and do not interact directly 
with the proteins~\cite{galinski,timasheff3,rszjk}. 

\begin{figure}[t!]
\includegraphics[width=8.5cm]{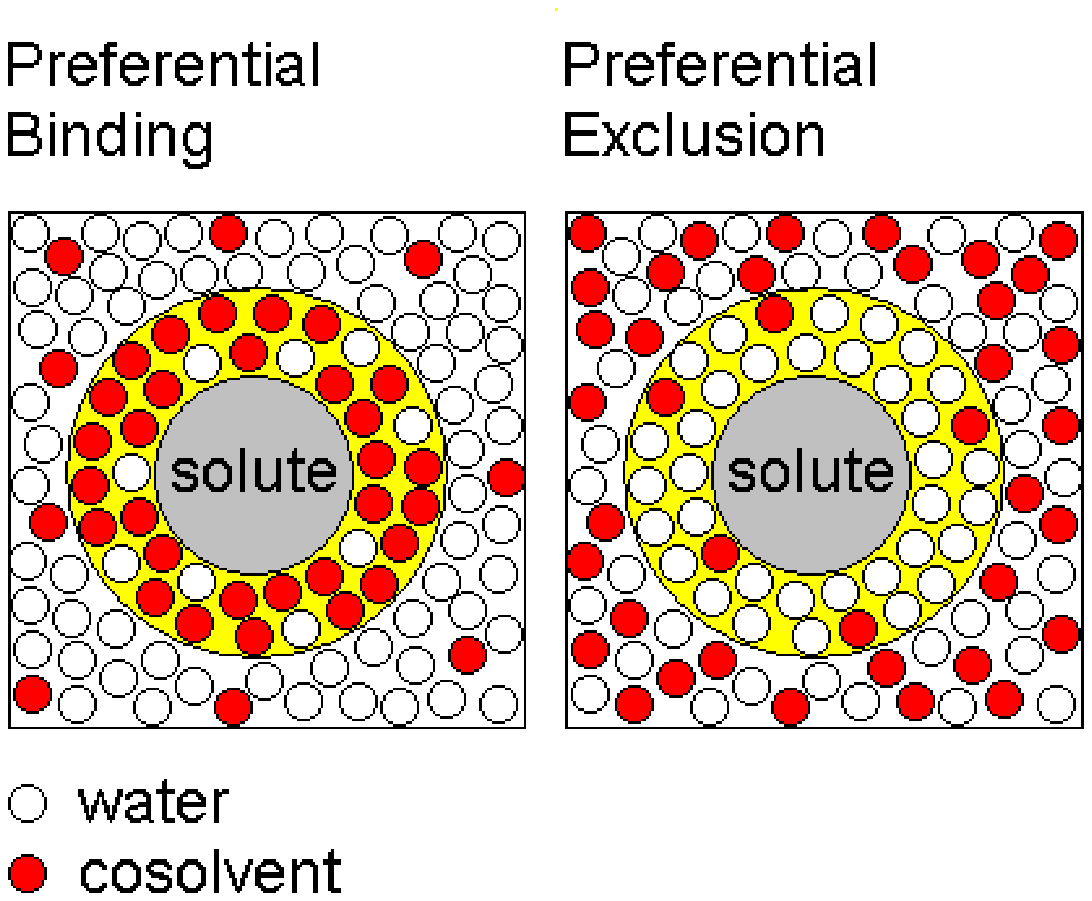}
\caption{Schematic representation of preferential phenomena in a mixture 
of water and hydrophobic solute particles in the presence of chaotropic 
(left) and kosmotropic (right) cosolvents (after Ref.~\cite{creighton}).}
\label{Fpref}
\end{figure}

Chaotropic cosolvents, such as urea in most ternary systems, are less polar 
than water, so that their presence in solution leads to an energetically 
unfavourable disruption of water structure. Such cosolvents are therefore 
excluded from bulk water, an effect known as ``preferential binding'' to 
the solute particles~\cite{timasheff2,creighton}, although it relies less 
on any direct binding of cosolvent to solute (which would enhance the 
effect) than on the fact that the cosolvent molecules are pushed from the 
solvent into the shell regions of the solute. Preferential binding and 
exclusion of cosolvent molecules are depicted in Fig.~\ref{Fpref}. In 
the former case, the smaller number of water molecules in contact with 
the surface of non-polar solute particles leads to a weaker effective 
interaction between solute and solvent, such that a larger mutual 
interface becomes favourable. The addition of chaotropic cosolvents to 
aqueous solutions therefore results in an increase in solvent-accessible 
surface area which destabilises hydrophobic aggregates, micelles and 
native protein structures~\cite{schellman,timasheff2,creighton,timasheff}. 

The essential phenomenology of the kosmotropic and chaotropic effects 
may be explained within a solvent-based model founded on the concept 
of two physically distinct types of solvent state, namely ordered and 
disordered water \cite{wiggins,galinski} (Fig.~\ref{Figws}). In 
Sec.~\ref{secModel} we review this model, described in detail in 
Ref.~\cite{moelbert2}, and focus on the adaptations which represent 
the structure-changing effects of the cosolvents. Sec.~\ref{secKosmo} 
presents the results of Monte Carlo simulations for kosmotropic 
substances, which illustrate their stabilising effect on hydrophobic 
aggregation, the decrease in solubility of non-polar molecules in 
water-cosolvent mixtures and the underlying preferential exclusion of 
cosolvent molecules from the solvation shell of hydrophobic particles. 
We consider the inverse phenomena of the chaotropic effect in 
Sec.~\ref{secChao}. In Sec.~\ref{secUNaCl} we illustrate these results 
by comparison with available experimental data for a kosmotropic and a 
chaotropic cosolvent, and discuss the extent of the symmetry between 
the two effects in the context of the model description for weakly or 
strongly active agents. Sec.~\ref{secConclusion} contains a summary 
and conclusions. 

\section{Model} 
\label{secModel}

\subsection{Hydrophobic-Polar Model}

The driving force in the process of solvation and aggregation is the 
effective hydrophobic interaction between polar water and the non-polar 
solute~\cite{kauzmann,franks}. As noted in Sec.~I, the origin of this 
interaction is the rearrangement of water around the solute particle. 
For solute surfaces of sufficiently high curvature to permit (partial) 
cage formation, this process decreases the enthalpy due to reinforced 
hydrogen bonds between water molecules in the solvation shell of the 
hydrophobic particle in comparison to those in bulk water, but reduces 
simultaneously the number of degenerate states of the solvent. These 
physical features are described by the model of Muller, Lee and Graziano 
(MLG), where the energy levels and respective degeneracies of water 
molecules are determined by the local water structure~\cite{muller,lee}. 

Because solute particles are relatively large compared with single water 
molecules, we use an adapted version of the MLG model in which each site 
contains a group of water molecules. The distinction between solvent 
molecules and non-polar solute particles lies in their ability to form 
hydrogen bonds. The continuous range of interaction energies within a 
partially hydrogen-bonded water cluster may be simplified to two discrete 
states of predominantly intact or broken hydrogen bonds~\cite{silverstein1}. 
The water sites in the coarse-grained model may then be characterised by 
two states, where an ``ordered'' site represents a water cluster with 
mostly intact hydrogen bonds, while a ``disordered'' site contains 
relatively fewer intact hydrogen bonds (Fig.~\ref{Figws}). The use of 
a bimodal distribution in the adapted model is not an approximation, 
but a natural consequence of the two-state nature of the pure MLG 
model~\cite{moelbert2}. In the presence of a non-polar solute a further 
distinction is required, between ``bulk'' water sites undisturbed by the 
solute particles and ``shell'' sites in their vicinity whose hydrogen 
bonding is altered. As explained in more detail below, this adaptation 
of the MLG model contains the essential features required to encapsulate 
the enthalpy/entropy balance which is the basis for the primary phenomena 
of hydrophobic aggregation in two-component water/solute mixtures. The 
model has been used to reproduce the appearance of upper (UCST) and lower 
(LCST) critical solution temperatures and a closed-loop coexistence 
regime~\cite{moelbert1}, demonstrating that the origin of the hydrophobic 
interaction lies in the alteration of water structure. 

The adapted MLG model has been extended to provide a minimal model for 
the study of chaotropic phenomena in ternary water/solute/cosolvent 
systems~\cite{moelbert2}. It was shown that this framework yields a 
successful description of preferential binding, and of the resulting 
destabilisation of aggregates of solute particles as a consequence of 
the role of chaotropic cosolvents in reducing the formation of water 
structure. We begin with a brief review of the model to summarise its 
physical basis and to explain the modifications required for the 
inclusion of cosolvent effects. 

\begin{figure}[t!]
\includegraphics[width=7.5cm]{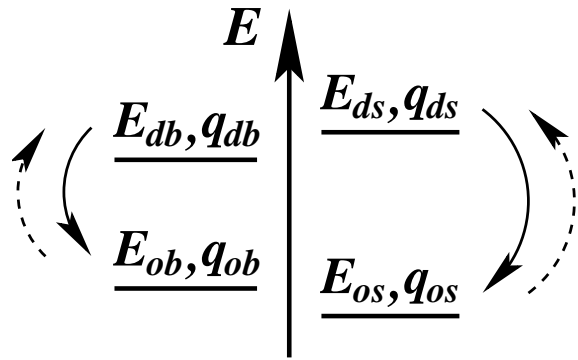}
\caption{Energy levels of a water site in the bimodal MLG model. The 
energy levels of shell water are different from bulk water due to breaking 
and rearrangement of hydrogen bonds in the proximity of a solute particle. 
The solid arrows represent the effect of a kosmotropic cosolvent in 
creating more ordered states, and the dashed arrows the opposite effect 
of chaotropic species. }
\label{FigMLG}
\end{figure}

The microscopic origin of the energy and degeneracy parameters for the 
four different types of water site (ordered/disordered and shell/bulk) is 
discussed in Refs.~\cite{moelbert2,silverstein2,paschek}, in conjunction 
with experimental and theoretical justification for the considerations 
involved. Here we provide only a qualitative explanation of their 
relative sizes, which ensure the competition of enthalpic and entropic 
contributions underlying the fact that the model describes a closed-loop 
aggregation region bounded by upper and lower critical temperatures 
and densities~\cite{moelbert1}. Considering first the energies 
of water sites (groups of molecules), the strongest hydrogen bonding 
arises not in ordered bulk water, but in the shell sites of hydrophobic 
particles due to the cage formation described in Sec.~I (which can be 
considered as a result of the orientational effect of a hydrophobic 
surface). Both types of ordered water cluster are enthalpically much 
more favourable than disordered groups of molecules, for which the 
presence of a non-polar particle serves only to reduce hydrogen bonding 
still further, making the disordered shell energy the highest of all. 
As regards the site degeneracies, which account for the entropy of 
the system, cage formation permits very few different molecular 
configurations, and has a low degeneracy. Water molecules in ordered 
clusters have significantly reduced rotational degrees of freedom compared 
to disordered ones, with the result that the degeneracy of the latter is 
considerably higher. Finally, the degeneracy of a disordered shell site 
is still higher than that of a bulk site because the relative reduction 
of hydrogen bonding due to the non-polar surface admits a higher amount 
of rotational freedom. 

Figure~\ref{FigMLG} summarises the energy levels of a water site, which 
are arranged in the sequence $E_{ds} > E_{db} > E_{ob} > E_{os}$. Their 
respective degeneracies are $q_{ds} > q_{db} > q_{ob} > q_{os}$, where 
the states are denoted $ds$ = disordered shell, $db$ = disordered bulk, 
$ob$ = ordered bulk and $os$ = ordered shell (cage conformation). We 
stress that these sequences are neither an assuption of the model nor 
may they be altered if the model is to represent a system exhibiting 
the closed-loop form of the coexistence curve for aggregation: only these 
specific sequences reproduce appropriately the competition of enthalpic 
and entropic contributions to the free energy. From the microscopic 
considerations of the previous paragraph these sequences are entirely 
plausible, and they have been confirmed by a range of experimental 
measurements; the exact values of the parameters are not important for 
the qualitative properties of the system while the order of energies and 
degeneracies is maintained. We return in Sec.~V to the issue of departure 
from these sequences. 

In the calculations to follow we have used the energy values $E_{ds} = 
1.8,\ E_{db} = 1.0,\ E_{ob} = -1.0$ and $E_{os} = -2.0$, which are 
thought to be suitably representative for aqueous solutions, and which 
have been successful in describing a number of different types of 
solution~\cite{delos2,silverstein2,moelbert1}. Their corresponding 
degeneracies, normalised to a non-degenerate ordered shell conformation, 
are taken as $q_{ds} = 49,\ q_{db} = 40,\ q_{ob} = 10$ and $q_{os} = 
1$~\cite{silverstein2}. We remind the reader that these energy and 
degeneracy parameters are independent of temperature and solute density.
Their relative values have been found to be appropriate for reproducing 
the primary qualitative features of hydrophobic 
interactions~\cite{silverstein2,moelbert1}, swelling of 
biopolymers~\cite{delos2}, protein denaturation~\cite{delos,caldarelli}, 
and micelle formation~\cite{rmnd}. The essential behaviour obtained 
within this model is indeed found to be rather insensitive to the 
precise parameter values, which may, however, be refined by 
comparison with experimental measurements to yield semi-quantitative 
agreement for different solutions~\cite{delos2,silverstein2,moelbert1}. 
The energy scale is correlated directly with a temperature scale, which 
we define as $k_B T\equiv \beta^{-1}$.

We conclude this subsection by qualifying the range of validity of the 
adapted MLG model, including what is meant above by ``small'' solute 
particles. The model is not appropriate for solute species smaller than 
a group of water molecules with intercluster hydrogen-bonding, which 
necessitates a linear cluster (and solute) size of at least 2 water 
molecules. The model will also fail to capture the physics of large, 
uniform systems where the linear cluster size exceeds perhaps 10 water 
molecules, because in this case the majority of the water molecules in 
a ``shell site'' would correspond effectively to bulk water. Most 
molecular-scale treatments of water structure in the vicinity of 
non-polar solute particles assume a spherical and atomically flat 
solute surface. These analyses \cite{rlmr,rhea,rsc} indicate that 
``clathrate'' models for the (partial) formation of hydrogen-bonded 
cages become inappropriate for larger curvatures, on the order of 1 nm, 
where a crossover occurs to ``depletion'' or ``dangling-bond'' models 
in which hydrogen-bond formation is frustrated \cite{rlcw}. While the 
adapted MLG model becomes inapplicable for such particles at the point 
where a clathrate description breaks down, we note \cite{rmnd} that most 
solute species, and in particular proteins, are atomically rough, in 
that they composed of chains and side-groups with high local curvatures 
on the order of 0.3-0.5 nm. As a consequence the model may remain valid 
for solute particles of considerably greater total dimension. For the 
purposes of the present analysis, we note that a significant quantity 
of the available data concerning cosolvent effects has been obtained 
for protein solutions, and that the majority of this data is consistent 
with our basic picture of the dominance of water structure effects in 
dictating solute solubility. However, while most proteins may indeed 
fall in the class of systems well described by a clathrate picture, 
we stress that the adapted MLG model cannot be expected to provide a 
complete description for large solute molecules containing many 
amino-acids of different local hydrophobicity and chemical interactions.

\subsection{Cosolvent Addition}

In comparison with the hydrophobic solute particles, cosolvent molecules 
are generally small and polar, and are therefore included directly in water 
sites by changing the number of states of these sites. A site containing 
water molecules and a cosolvent particle is referred to as a cosolvent site. 
Kosmotropic cosolvents, being more polar than water, increase the number 
of intact hydrogen bonds at a site~\cite{walrafen}. In the bimodal MLG 
framework their addition to weakly hydrogen-bonded, disordered clusters may be 
considered to create ordered clusters with additional intact hydrogen bonds. 
The creation of ordered states from disordered ones in the presence of a 
kosmotropic cosolvent increases the degeneracy of the former at the 
expense of the latter. This feature is incorporated by raising the number 
of possible ordered states compared to the number of disordered states 
(solid arrows in Fig.~\ref{FigMLG}), 
\begin{eqnarray} \nonumber
q_{ob, k} =  q_{ob} + \eta_b, &\hspace{1cm} & q_{db, k} = q_{db} - \eta_b, \\
q_{os, k} =  q_{os} + \eta_s, &\hspace{1cm} & q_{ds, k} = q_{ds} - \eta_s, 
\label{eqDegen}
\end{eqnarray}
where $k$ denotes the states of water clusters containing kosmotropic 
cosolvent molecules, and the total number of states is kept constant. 

The effect of chaotropic cosolvents on the state degeneracies is opposite 
to that of kosmotropic substances, and may be represented by inverting 
the signs in eq.~\ref{eqDegen}. Chaotropic cosolvents are less strongly 
polar than water, acting in an aqueous solution of hydrophobic particles 
to reduce the extent of hydrogen bonding between water molecules in both 
shell and bulk sites~\cite{franks,finer}. Within the adapted MLG framework, 
this effect is reproduced by the creation of disordered states, with 
additional broken hydrogen bonds and higher enthalpy, from the more 
strongly bonded ordered clusters (dashed arrows in Fig.~\ref{FigMLG}), 
whence
\begin{eqnarray} \nonumber
q_{ob, c} =  q_{ob} - \eta_b, &\hspace{1cm} & q_{db, c} = q_{db} + \eta_b, \\
q_{os, c} =  q_{os} - \eta_s, &\hspace{1cm} & q_{ds, c} = q_{ds} + \eta_s.
\label{eqDegen2}
\end{eqnarray}

We stress three important qualitative points. First, the cosolvent 
affects only the number of intact hydrogen bonds, but not their strength 
\cite{franks}, as a result of which, also in the bimodal distribution of 
the coarse-grained model, the energies of the states remain unchanged. 
Cosolvent effects are reproduced only by the changes they cause in the 
relative numbers of each type of site [eqs.~(\ref{eqDegen}) and 
(\ref{eqDegen2})]. Secondly, cosolvent effects on the relative 
degeneracy parameters are significantly stronger in the bulk than 
in the shell, because for obvious geometrical reasons related to 
the orientation of water molecules around a non-polar solute 
particle~\cite{moelbert2,silverstein2}, many more hydrogen-bonded 
configurations are possible in the bulk. Finally, as suggested in 
Sec.~I, our general framework does not include the possibility of 
cosolvent-solute interactions, which have been found to be important in 
certain ternary systems, specifically those containing urea \cite{rb,rkr}.

The illustrative calculations in Secs.~III and IV are performed with 
$\eta_b = 9.0$ and $\eta_s = 0.1$ in eqs.~(\ref{eqDegen}) and 
(\ref{eqDegen2}). In the analysis of Ref.~\cite{moelbert2} these 
values were found to provide a good account of the qualitative physics 
of urea as a chaotropic cosolvent; in Sec.~V we will return to a more 
quantitative discussion of the role of these parameters. By using the 
same values for a hypothetical kosmotropic cosolvent we will demonstrate 
that a symmetry of model degeneracy parameters does not extend to a 
quantitative, or even qualitative, symmetry in all of the relevant 
physical phenomena caused by structure-changing cosolvents. We comment 
briefly that the fractional value of $\eta_s$ arises only from the 
normalisation convention $q_{os} = 1$. 

\subsection{Mathematical Formulation}

To formulate a description of the ternary solution within the framework of 
statistical mechanics, we begin by associating with each of the $N$ sites 
of the system, labelled $i$, a variable $n_i$, which takes the values $n_i 
= 1$ if the site contains either pure water, $n_i = 0$  if the site 
contains a hydrophobic particle, or $n_i = -1$ if the site contains a 
water cluster including one or more cosolvent molecules. This system 
is described by the Hamiltonian 
\begin{eqnarray}\nonumber
&&\!\!\!\!\!\!\!H[\{n_i\},\{\sigma_i\}]\!=\! \sum_{i=1}^N  \frac{n_i 
(n_i\!+\!1)}{2}[(E_{ob} \tilde\delta_{i, \sigma_{ob}}\!+\! E_{db} 
\tilde\delta_{i, \sigma_{db}}) \lambda_i \\ \nonumber
&&+ (E_{os} \tilde\delta_{i, \sigma_{os}} + E_{ds} \tilde\delta_{i, 
\sigma_{ds}})(1-\lambda_i)]  \\ \nonumber
&&+ \sum_{i=1}^N \frac{n_i (n_i-1)}{2}[(E_{ob} \tilde\delta_{i, 
\sigma_{ob,a}} + E_{db} \tilde\delta_{i, \sigma_{db,a}}) \lambda_i \\ 
&&+ (E_{os} \tilde\delta_{i, \sigma_{os,a}} + E_{ds} \tilde\delta_{i, 
\sigma_{ds,a}})(1-\lambda_i)] ,
\label{eq:H}
\end{eqnarray}
where $a = k,c$ denotes sites containing water and a kosmotropic or 
chaotropic cosolvent molecule. The site variable $\lambda_i$ is defined 
as the product of the nearest neighbours, $\lambda_i = \prod_{\langle i,j 
\rangle} n_j^2$, and takes the value $1$ if site $i$ is completely 
surrounded by water and cosolvent or $0$ otherwise. The first sum 
defines the energy of pure water sites and the second the energy 
of cosolvent sites. Because a water site $i$ may be in one of $q$ 
different states, $\tilde\delta_{i, \sigma_{os}}$ is $1$ if site 
$i$ is occupied by water in one of the $q_{os}$ ordered shell states 
and $0$ otherwise, and $\tilde\delta_{i, \sigma_{ds}}$ is $1$ if it is 
occupied by pure water in one of the $q_{ds}$ disordered shell states 
and $0$ otherwise. Analogous considerations apply for the bulk states 
and for the states of cosolvent sites. A detailed description of the 
mathematical structure of the model is presented in Refs.~\cite{moelbert1} 
and \cite{moelbert2}.

The canonical partition function of the ternary $N$-site system, 
obtained from the sum over the state configurations $\{\sigma_i\}$, is 
\begin{eqnarray}\nonumber
Z_N &= \sum_{\{n_i\}} \prod_i &Z_{s}^{\frac{n_i (n_i+1)}{2}(1- \lambda_i)} 
Z_{b}^{\frac{n_i (n_i+1)}{2} \lambda_i}\\
&&\!\!\!\!\!\!\times \,Z_{s, a}^{\frac{n_i (n_i-1)}{2}(1- \lambda_i)} 
Z_{b, a}^{\frac{n_i (n_i-1)}{2} \lambda_i},
\end{eqnarray}
where $Z_{\sigma} = q_{o\sigma} e^{-\beta E_{o\sigma}} + q_{d\sigma} 
e^{-\beta E_{d\sigma}}$ for the shell ($\sigma = s$) and bulk ($\sigma 
= b$) states both of pure-water and of cosolvent sites ($\sigma = s,a$ 
and $\sigma = b,a$).

When the number of particles may vary, a chemical potential is associated 
with the energy of particle addition or removal. The grand canonical 
partition function of the system for variable particle number becomes 
\begin{equation} 
\Xi = \sum_{N} e^{\beta \mu N_w + \beta (\mu + \Delta \mu) N_a} Z_N = 
\sum_{\{n_i\}} e^{-\beta H_{\rm eff}^{\rm gc}[\{n_i\}]} .
\label{eq:gcZkosmo}
\end{equation} 
$N_w$ denotes the number of pure water sites, $N_a$ the number of 
cosolvent sites and $N_p$ the number of solute particle sites, the 
total number being $N = N_w + N_a + N_p$. The variable $\mu$ represents 
the chemical potential associated with the addition of a water site to 
the system and $\Delta \mu$ the chemical potential for the insertion of 
a cosolvent molecule at a water site. From the role of the two cosolvent 
types in enhancing or disrupting water structure, at constant solute 
particle density $\Delta \mu$ is expected to be negative for kosmotropic 
agents and positive for chaotropic ones. 

\subsection{Monte Carlo Simulations}

We describe only briefly the methods by which the model may be analyzed; 
full technical details may be found in Ref.~\cite{moelbert1}. Because our 
primary interest is in the local solute and cosolvent density variations 
which demonstrate aggregation and preferential phenomena, we focus in 
particular on Monte Carlo simulations at the level of individual solute 
molecules. Classical Monte Carlo simulations allow the efficient 
calculation of thermal averages in many-particle systems with 
statistical fluctuations~\cite{moelbert1}, such as that represented by 
eq.~(\ref{eq:gcZkosmo}). We work on a cubic lattice of $30\times30\times30$ 
sites ($N = 27\,000$), with random initial particle distributions and with 
periodic boundary conditions to eliminate edge effects (although clearly 
we cannot eliminate finite-size effects). Each site is occupied 
by either a solute particle, pure water or a water-cosolvent mixture. The 
results are unaffected by changes in lattice size. We have implemented a 
Metropolis algorithm for sampling of the configuration space. After a 
sufficiently large number ($100\,000$) of relaxation steps, the system 
achieves thermal equilibrium and averages are taken over a further 
$500\,000$ measurements to estimate thermodynamic quantities. 
Coexistence lines in the $\mu$-$T$ phase diagram are obtained from the 
transitions determined by increasing the temperature at fixed chemical 
potential $\mu$ (grand canonical sampling), which results in a sudden 
density jump at the transition temperature. The solute particle densities 
$\rho_p$ corresponding to these jumps yield closed-loop coexistence curves 
in the $\rho_p$-$T$ phase diagram. Despite the rather crude approximation 
to a continuum system offered by the use of a cubic lattice, we have found 
quantitative agreement with other theoretical approaches, which is why we 
focus primarily on Monte Carlo simulations here. In Sec.~V we will also 
demonstrate remarkable qualitative, and in some cases semi-quantitative, 
agreement with experiment, suggesting that the adapted MLG framework may 
indeed form a suitable basis for more sophisticated modelling.

For the macroscopic properties of the system, such as aggregation and phase
coexistence, the results of the Monte Carlo simulations may be interpreted 
with the aid of a mean-field analysis. If the densities at each site are 
approximated by their average values in the solution, $\rho_p$ for 
hydrophobic particles, $\rho_w$ for pure water and $\rho_a$ for cosolvent, 
where $\rho_p + \rho_w + \rho_a = 1$, the mean value $\langle n_i\rangle$ 
at site $i$ is given by $\langle n_i\rangle = \sum_{\sigma }n_{i,\sigma} 
\rho_{\sigma} = \rho_w - \rho_a$ and analogously $\langle n_i^2 \rangle 
= \sum_{\sigma} n^2_{i,\sigma} \rho_{\sigma}  =  \rho_w + \rho_a$. The 
equilibrium densities are obtained by minimisation of the grand canonical 
mean-field free energy per site, 
\begin{eqnarray}\nonumber
f & = & (\mu - \beta^{-1} \ln Z_{s}) \rho_w + (\mu + \Delta \mu - 
\beta^{-1}  \ln Z_{s,a}) \rho_a \\ \nonumber
& & + \beta^{-1} (\ln Z_{s}-\ln Z_{b}) \rho_w (\rho_a+\rho_w)^{z} \\ \nonumber
& & + \beta^{-1} (\ln Z_{s,a}-\ln Z_{b,a}) \rho_a (\rho_a+\rho_w)^{z} \\
& & + \beta^{-1} (\rho_w \ln \rho_w + \rho_a \ln \rho_a +\rho_p \ln \rho_p),
\label{eq:fkosmo}
\end{eqnarray}
where $z$ denotes the number of nearest neighbours of each site. 

\section{Kosmotropic Effect}
\label{secKosmo}

\begin{figure}[t!]
\includegraphics[width=8.5cm]{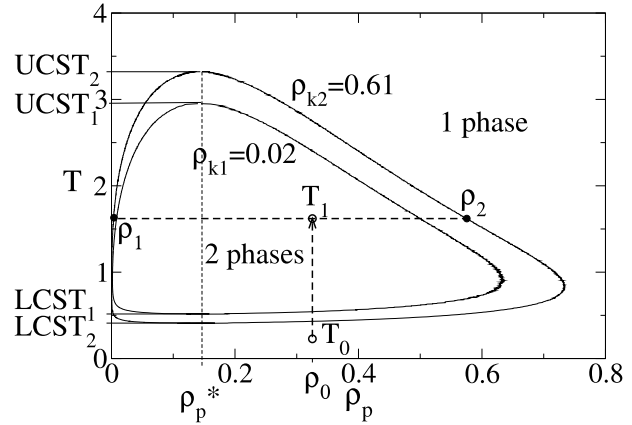}
\caption{Closed-loop coexistence curves for a ternary system of water, 
kosmotropic cosolvent and hydrophobic particles for two different cosolvent 
densities $\rho_k$, obtained by mean-field calculation with degeneracy 
parameters $\eta_b = 9.0$ and $\eta_s = 0.1$ in eq.~(\ref{eqDegen}). On the 
outside of each curve the solution is homogeneous, whereas on the inside 
it separates into two phases. The dashed arrow represents the heating 
process of a system with particle density $\rho_0$ and cosolvent density 
$\rho_{k2} = 0.61$ from temperature $T_0$ in the homogeneous region to 
$T_1$ in the two-phase region. At $T_1$ the system is separated into two 
phases of different densities $\rho_1$ (nearly pure water-cosolvent 
mixture) and $\rho_2$ (hydrophobic aggregates).  }
\label{FigBouleMFKosmo}
\end{figure}

We begin our analysis of the kosmotropic effect by discussing the 
mean-field $\rho_p$-$T$ phase diagram for different cosolvent 
concentrations, shown in Fig.~\ref{FigBouleMFKosmo}. As expected 
from Sec.~IIA the system exhibits a closed-loop coexistence curve,
forming a homogeneous particle-solvent-cosolvent mixture below a LCST 
and above an UCST for all concentrations, whereas between these 
temperatures, and for intermediate particle concentrations, a phase 
separation is found into a pure solvent phase (meaning a water-cosolvent 
mixture) and an aggregated phase with fixed solute density. 

The hydrophobic repulsion between water and non-polar solute particles, 
developed as a result of the formation of water structure, increases in 
the presence of kosmotropic cosolvents (this result may be shown by 
rewriting the partition function (eq.~\ref{eq:gcZkosmo}) in terms of the 
particle sites~\cite{moelbert2}). As shown in Fig.~\ref{FigBouleMFKosmo}, 
the temperature and density ranges of the aggregation region indeed rise 
with increasing cosolvent concentration, illustrating the stabilising 
effect of kosmotropic agents. Further, the particle density of the 
aggregate phase in the two-phase region increases when adding cosolvent, 
demonstrating in addition the strengthening of effective hydrophobic 
interactions between solute particles due to the growing water-solute 
repulsion. This is illustrated by the process of heating a system at 
constant density: the dashed arrow in Fig.~\ref{FigBouleMFKosmo} represents 
a solution with particle density $\rho_0$, which is heated from temperature 
$T_0$ in the homogeneous region to a temperature between the LCST and the 
UCST. In the heterogeneous region ($T_1$), the solution separates into 
two phases of densities $\rho_1$ (almost pure solvent) and $\rho_2$ 
(hydrophobic aggregates) under the constraint $\rho_0 (V_1 + V_2) = 
\rho_1 V_1 + \rho_2 V_2$, where $V_i$ is the volume occupied by phase 
$i$. An increase in cosolvent density leads to a higher density $\rho_2$ 
of hydrophobic aggregates and a lower value $\rho_1$, which may be 
interpreted as a strengthening of the hydrophobic interactions 
generated between solute particles. 

\begin{figure}[t!]
\includegraphics[width=8.5cm]{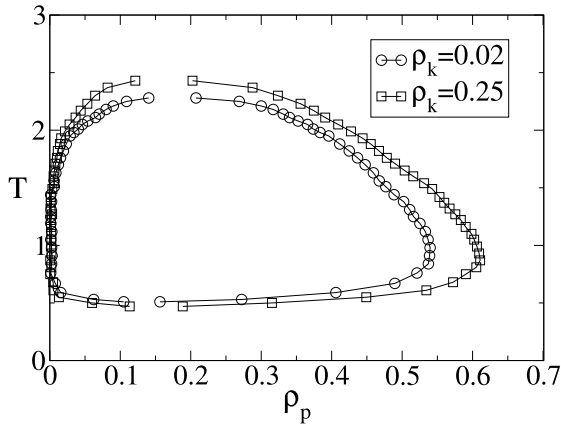}
\caption{Closed-loop coexistence curves obtained by Monte Carlo 
simulations for an aqueous solution of hydrophobic particles with the 
same parameters as in Fig.~\ref{FigBouleMFKosmo}. }
\label{FigBouleMCKosmo}
\end{figure}

However, the mean-field approximation neglects fluctuations in the density, 
and consequently is unable to detect intrinsically local phenomena such as 
preferential exclusion of cosolvent particles from the solvation shell of a 
hydrophobic particle (Fig.~\ref{Fpref}). Because the kosmotropic effect is 
strongly dependent upon local enhancement of water structure, a full 
analysis is required of the spatial density fluctuations in the system. 
We thus turn to the results of Monte Carlo simulations, which at the 
macroscopic level show the same enhancement of hydrophobic aggregation 
(Fig.~\ref{FigBouleMCKosmo}) as indicated in the mean-field analysis. 
Below the LCST and above the UCST, one homogeneous mixture is found where 
the hydrophobic particles are dissolved, while between these temperatures 
they form aggregates of a given density. The LCST and the critical aggregate 
densities determined by the mean-field calculation do not differ strongly 
from the numerical results, but the UCST lies at a temperature higher than 
that determined by Monte Carlo simulations. This is to be expected 
because mean-field calculations neglect fluctuation effects, generally 
overestimating both transition temperatures. The agreement of the 
mean-field result for small $\rho_k$ with the simulated value for the 
LCST suggests a predominance of local effects at low temperatures, and 
that density fluctuations between sites are small. We note, however, that 
a cosolvent concentration $\rho_k = 0.61$ is required in the mean-field 
approximation to produce an expansion of the aggregated phase similar to 
that observed in the Monte Carlo results for $\rho_k = 0.25$, demonstrating 
the growing importance of fluctuation effects at higher cosolvent 
concentrations.

\begin{figure}[t!]
\includegraphics[width=8.5cm]{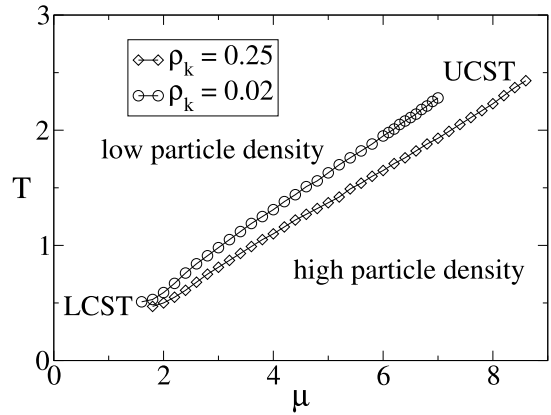}
\caption{$\mu$-$T$ phase diagram of hydrophobic solute in an aqueous 
cosolvent mixture for two different kosmotropic cosolvent concentrations 
$\rho_k$, obtained by Monte Carlo simulations. The endpoints of the 
finite transition lines correspond to the UCST and LCST. }
\label{FigMuTKosmo}
\end{figure}

The $\mu$-$T$ phase diagram of the ternary system obtained by Monte 
Carlo simulations is presented in Fig.~\ref{FigMuTKosmo}. Lines of 
finite length terminated by the UCST and LCST demarcate the aggregation 
phase transition. An increase in cosolvent density leads to a larger 
separation of UCST and LCST as a direct result of the expansion of 
the aggregation regime (Fig.~\ref{FigBouleMCKosmo}). The transition 
line is shifted to higher values of $\mu$, indicating an increased 
resistance of the system to addition of water at constant volume in 
the presence of cosolvent. This is a consequence of the fact that the 
resistance to addition of hydrophobic particles depends on both $\mu$ 
and $\Delta\mu$: with increasing cosolvent concentration, $\Delta\mu$ 
decreases ({\it i.e.}~becomes more negative) and hence $\mu$ must 
increase for the critical particle density to remain constant. 

By inspection of the energy levels, and of the alteration in their relative 
degeneracies caused by a kosmotropic cosolvent (Fig.~\ref{FigMLG}), 
it is energetically favourable to maximise the cosolvent concentration
in bulk water with respect to shell water. Thus one expects a 
preferential exclusion, also known as preferential hydration, as
represented schematically in the right panel of Fig.~\ref{Fpref}. 
Except at the lowest temperatures, an increased number of water 
molecules in the solvation shell of non-polar solute particles 
increases the repulsion between solute and solvent, resulting in a 
reduction of solubility, in an enhanced effective attraction between 
solute particles due to the decreased interface with water as they 
approach each other, and thus in aggregate stabilisation. 

\begin{figure}[t!]
\includegraphics[width=8.5cm]{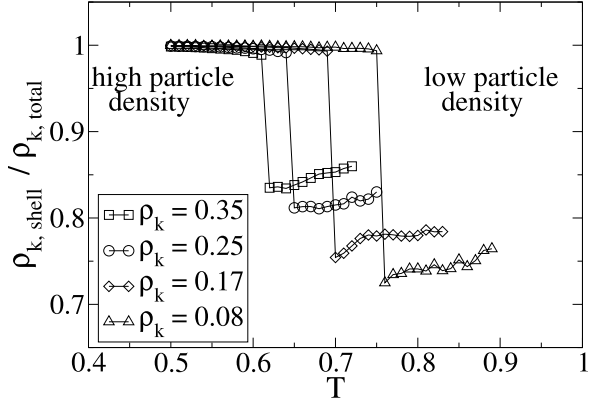}
\caption{Relative concentration of kosmotropic cosolvent in the 
solvation shell of hydrophobic particles, obtained from Monte Carlo 
simulations at chemical potential $\mu = 2.5$, exhibiting preferential 
exclusion. The effect is most pronounced for low cosolvent densities.  }
\label{FigShellBulkKosmo}
\end{figure}

Monte Carlo simulations confirm the expectation that the 
cosolvent concentration is lower in shell water than in bulk water. 
Figure~\ref{FigShellBulkKosmo} shows the cosolvent shell concentration 
compared to the overall cosolvent concentration in the solution, where 
preferential exclusion of the kosmotropic cosolvent from the solvation 
shell of the hydrophobic solute particles is observed. At constant chemical 
potential $\mu$, a sharp drop occurs at the phase transition temperature 
(Fig.~\ref{FigMuTKosmo}), which is due to the sudden change in solute 
particle density. At very low particle density, a clear exclusion of 
the cosolvent from the solvation shell appears. However, at high 
particle density (for temperatures below the aggregation phase 
transition) the effect is only marginal, because here the number 
of particle sites becomes substantial, most solvent and cosolvent 
sites are shell sites and thus the total cosolvent density is almost 
identical to the shell cosolvent density. The comparatively strong 
fluctuations in relative cosolvent shell concentration at temperatures 
above the phase transition may be attributed to the very low particle 
density in the system. Small, thermal fluctuations in the number of 
shell sites occupied by cosolvent molecules result in large fluctuations 
in the cosolvent shell concentration. 

Considering next the effect of the cosolvent concentration, the system 
shows a substantially stronger preferential exclusion for low than for 
high cosolvent densities. This is again largely a consequence of 
considering the relative, as opposed to absolute, decrease in cosolvent 
concentration in the shell. Preferential exclusion is less pronounced 
for high cosolvent densities because a smaller proportion of cosolvent 
sites can contribute to the effect. At high temperatures, entropy, and 
therefore mixing, effects become dominant and the preferential exclusion 
of cosolvent particles shows a slight decrease. 

We conclude this section by commenting also on the phenomenon shown 
in Figs.~\ref{FigBouleMFKosmo} and \ref{FigBouleMCKosmo}, that for 
temperatures where no aggregation is possible in the pure water-solute 
solution (below the LCST), it can be induced only by the addition of a 
cosolvent. Aggregation is driven by exclusion of solute from the 
solution in this intermediate regime, and is thus an indirect result of 
the transformation of disordered bulk water states into energetically 
favourable ordered bulk states by the kosmotropic cosolvent. Preferential 
hydration is a further manifestation of this exclusion. 

\section{Chaotropic Effect}
\label{secChao}

\begin{figure}[t!]
\includegraphics[width=8cm]{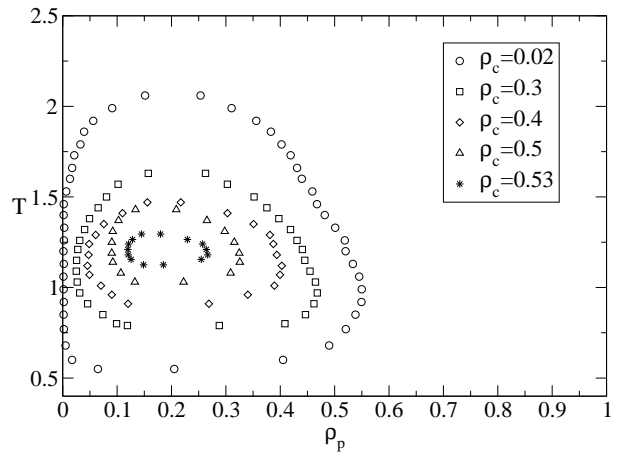}
\caption{Closed-loop miscibility curves of an aqueous solution of 
hydrophobic particles for different chaotropic cosolvent concentrations 
$\rho_c$, obtained by Monte Carlo simulations using the degeneracy 
parameters $\eta_b = 9.0$ and $\eta_s = 0.1$ in eq.~(\ref{eqDegen2}). 
On the outside of the curves is the homogeneous, dissolved state while 
inside them is the aggregated state. As the cosolvent concentration 
grows, particle solubility increases, leading to a reduced coexistence 
regime. }
\label{FigBouleMC_chaot}
\end{figure}

A detailed study of chaotropic cosolvent effects within the 
adapted MLG model for a hydrophobic-polar mixture was presented in 
Ref.~\cite{moelbert2}. Here we provide only a brief review of the results 
of this analysis, with emphasis on the differences and similarities in 
comparison with the kosmotropic effect, which we have discussed at 
greater length in Sec.~\ref{secKosmo}. Figure~\ref{FigBouleMC_chaot} 
shows the coexistence curves demarcating the aggregated state for a 
range of cosolvent concentrations, computed by Monte Carlo simulations. 
The general features of the phase diagram are those of Sec.~\ref{secKosmo} 
(Figs.~\ref{FigBouleMFKosmo} and \ref{FigBouleMCKosmo}), with the 
obvious exception of the shrinking of the coexistence regime as 
cosolvent concentration is increased. However, it appears that for the 
$q$ and $\eta$ parameters used in the analysis (Sec.~\ref{secModel}B) 
the effect of high chaotropic cosolvent concentrations is rather 
stronger than that obtained with the kosmotropic cosolvent, an issue 
to which we return in Sec.~\ref{secUNaCl}. Figure \ref{FigBouleMC_chaot} 
suggests that sufficiently high concentrations of chaotropic cosolvent 
may cause complete aggregate destabilisation at any temperature and 
solute particle density. 

\begin{figure}[t!]
\includegraphics[width=8cm]{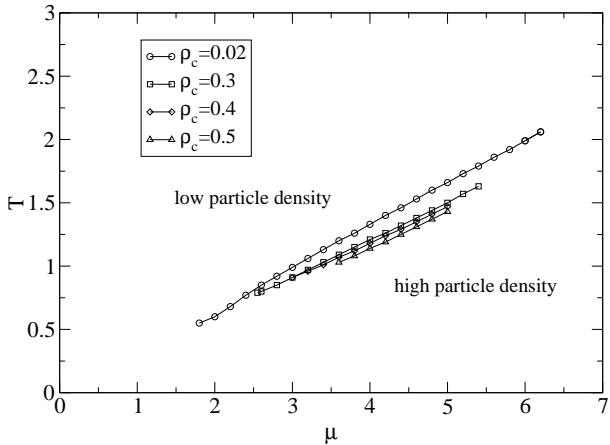}
\caption{$\mu$-$T$ phase diagram of hydrophobic solute in an aqueous 
cosolvent mixture for different chaotropic cosolvent concentrations $\rho_c$, 
obtained by Monte Carlo simulation. The finite transition lines terminate 
at an UCST and a LCST, which approach each other with increasing cosolvent 
concentration.  }
\label{FigMuTMC_chaot}
\end{figure}

Figure~\ref{FigMuTMC_chaot} shows the corresponding $\mu$-$T$ phase 
diagram. As in Fig.~\ref{FigMuTKosmo}, the lines representing the 
discontinuous jump in density from the low- to the high-$\rho_p$ phase 
move to higher $\mu$ with increasing cosolvent concentration, again 
reflecting the increased cost of adding water to the system in the 
presence of cosolvent. However, in this case the endpoints of the lines, 
which represent the UCST and LCST for each cosolvent concentration, 
become closer as $\rho_c$ increases until they approach a singularity 
(Fig.~\ref{FigBouleMC_chaot}), which we discuss in detail in 
Sec.~\ref{secUNaCl}.

\begin{figure}[t!]
\medskip
\includegraphics[width=8cm]{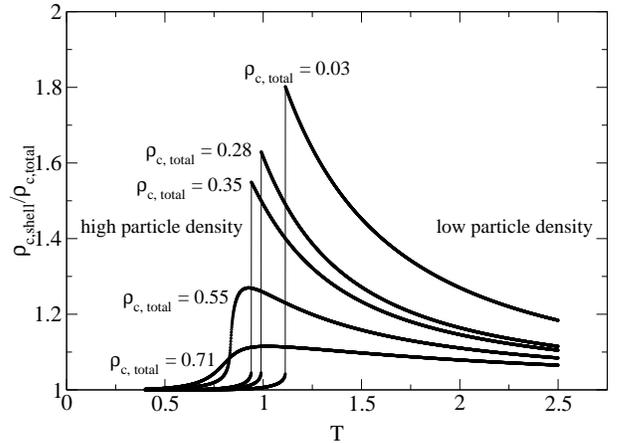}
\caption{Preferential binding effect, illustrated by the relative 
concentration of chaotropic cosolvent in the solvation shell of 
hydrophobic particles, for different cosolvent concentrations at 
fixed $\mu = 3.0$, obtained by pair approximation. }
\label{FigCrowdMuPA_chaot}
\end{figure}

Figure~\ref{FigCrowdMuPA_chaot} shows as a function of temperature the 
cosolvent concentration in the shell of a solute particle normalised by 
the total $\rho_c$. This demonstration of preferential binding of 
chaotropic agents to non-polar solute particles was obtained from the 
model of eq.~(\ref{eq:H}) within a pair approximation, which was shown 
in Ref.~\cite{moelbert2} to provide a semi-quantitative reproduction of 
the Monte Carlo results. As in Fig.~\ref{FigShellBulkKosmo}, the sharp 
step at the phase transition is due to the sudden change in particle 
density. At low particle densities (below the transition) the effect 
is clear, whereas at high densities it is small because the majority 
of solvent and cosolvent sites have become shell sites. For low 
cosolvent concentrations, preferential binding can lead to a strong 
increase in $\rho_c$ on the shell sites (80\% in 
Fig.~\ref{FigCrowdMuPA_chaot}), but the relative enhancement becomes 
less significant with increasing total cosolvent concentration. Finally, 
in counterpoint to the discussion at the end of Sec.~\ref{secKosmo}, we 
comment that for chaotropic agents, which in bulk water transform ordered 
states into disordered ones, preferential binding (cosolvent exclusion) 
is favoured and solute exclusion decreases. Thus a regime exists in the 
$\rho_p$-$T$ phase diagram where at fixed temperature and particle 
density, aggregates of solute particles may be made to dissolve simply 
by the addition of cosolvent.

\section{Strong and Weak Cosolvents}
\label{secUNaCl}

As indicated in Sec.~\ref{secModel} and illustrated by the results of 
Secs.~\ref{secKosmo} and \ref{secChao}, there is a qualitative symmetry 
between the opposing effects of kosmotropic and chaotropic substances
on hydrophobic aggregation of a given solute species. Indeed, the 
competition of the two effects has been studied for solutions of the 
protein HLL with the (denaturing) chaotrope guanidine hydrochloride and 
the (native-structure stabilising) kosmotrope betaine \cite{rszjk}. We 
remark, however, that the same cosolvent may have a different effect for 
solutions of different types and sizes of solute (Sec.~II), and for this 
reason a global characterisation of the effectiveness of any one agent 
may be misleading. 

At low cosolvent concentration, the stabilising or destabilising effects 
of the two cosolvent types on aggregate formation are rather small, and 
appear almost symmetrical in the expansion or contraction of the 
coexistence region (Figs.~\ref{FigBouleMCKosmo} and \ref{FigBouleMC_chaot}, 
Figs.~\ref{FigMuTKosmo} and \ref{FigMuTMC_chaot}). For weakly 
kosmotropic and chaotropic agents, by which is meant those causing 
only small enhancement or suppression of aggregation even at high 
concentrations, such qualitative symmetry might be expected as a general 
feature of the coexistence curves. With the same caveats concerning low 
concentrations or weak cosolvent activity, this statement is also true 
for preferential exclusion and binding of cosolvent to the solute 
particles, as shown by comparing Figs.~\ref{FigShellBulkKosmo} and 
\ref{FigCrowdMuPA_chaot}. However, differing phenomena emerge for 
strongly kosmotropic and chaotropic agents.

\begin{figure}[t!]
\centerline{\includegraphics[width=7.5cm]{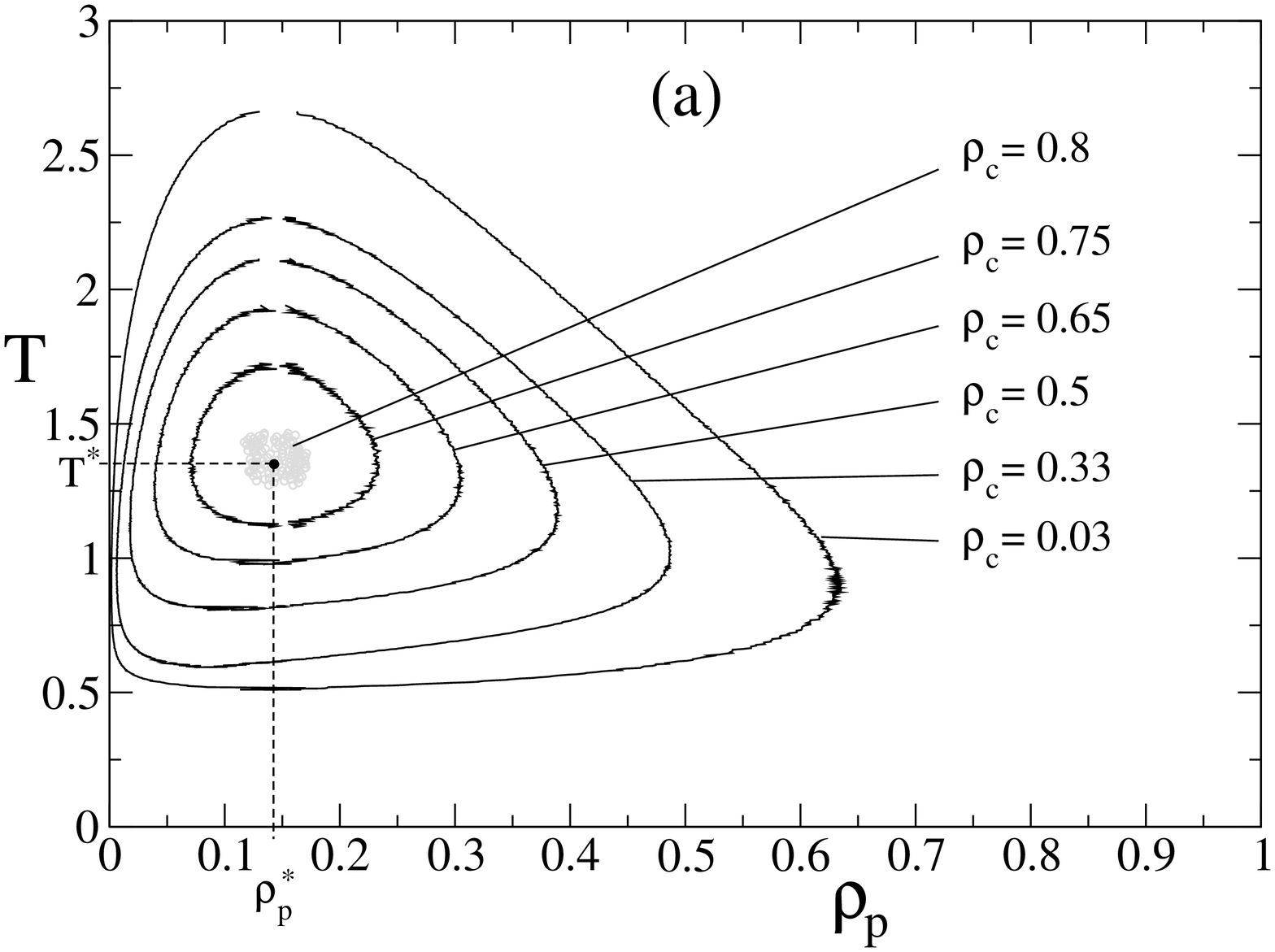}}
\vspace{1.0cm}
\centerline{\includegraphics[width=7.5cm]{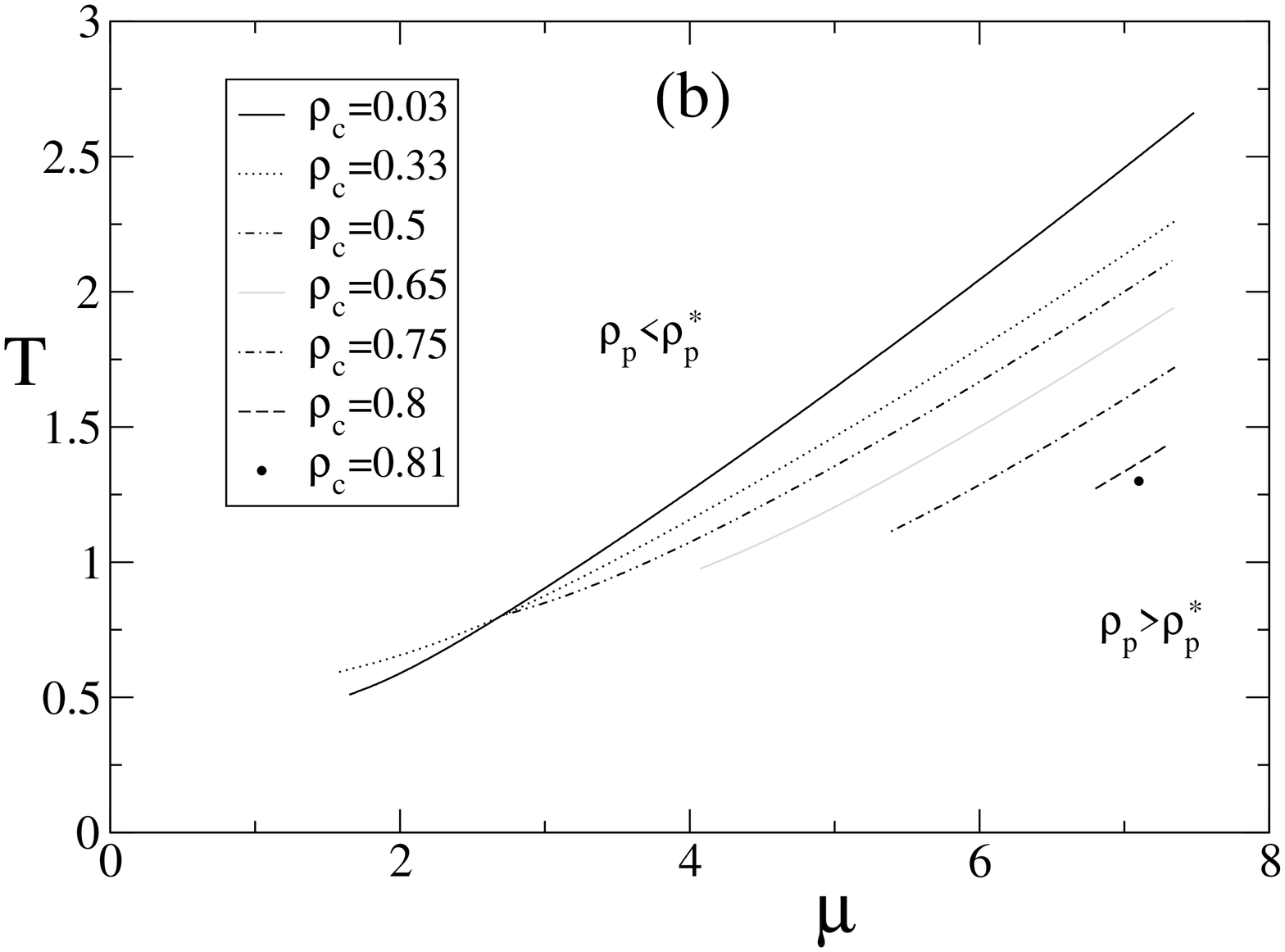}}
\caption{(a) Coexistence curves for ternary water/cosol- vent/solute 
systems with different cosolvent concentrations $\rho_c$, obtained 
by mean-field calculation. At a critical cosolvent concentration of 
$\rho_c^{\ast} = 0.81$, the LCST and UCST coincide, at a critical 
temperature $T^{\ast} = 1.35$ and critical particle density 
$\rho_p^{\ast} = 0.14$, where the aggregation regime is reduced to 
a single point. For still higher cosolvent concentrations the solution 
is homogeneous over the full range of temperature and particle 
density. (b) Mean-field $\mu$-$T$ phase diagram for different 
cosolvent concentrations. As $\rho_c$ increases, the separation of 
LCST and UCST decreases until for $\rho_c = \rho_c^{\ast}$ they meet 
at the temperature $T_L^{\ast} = T_U^{\ast} = T^{\ast}$, where the 
value $\mu_L = \mu_U = 7.05$ determines $\rho_p^{\ast} = 0.14$.}
\label{FigBoulemu_T_MF_chaot}
\end{figure}

\subsection{Urea as a Strong Chaotrope}

Chaotropic cosolvents act to suppress the formation of water 
structure and, because the latter is already disrupted by the presence 
of a non-polar solute particle, are more efficient for bulk than for 
shell sites. Within the adapted MLG model, a maximally chaotropic agent 
is one which causes sufficient disruption that the number of ordered 
bulk states is reduced to its hypothetical minimum value, which is 
equal to the number of ordered shell states; by the definition based 
on hydrogen-bond formation, it is not possible to have fewer ordered 
bulk than ordered shell states, so $q_{ob} \ge q_{os}$. When 
$q_{ob}$ and $q_{os}$ become similar, the entropic gain of creating 
ordered bulk states by aggregation (removal of shell sites) is largely 
excluded, and thus a strongly chaotropic agent raises the solubility 
of hydrophobic particles very significantly. With a sufficient 
concentration of such cosolvents, aggregation may be completely 
prevented at all temperatures and densities, and the coexistence regime 
vanishes at a critical value of cosolvent concentration. 

Such singular behaviour is well known in many aqueous solutions containing 
urea, and indeed frequent use is made of the destabilising properties 
of this cosolvent (Sec.~I). We stress at this point that the chaotropic 
action of urea is by no means universal: in a number of systems, such 
as the ultra-small solute methane \cite{wetlaufer,rg}, and proteins with 
variable hydrophobicity \cite{rsc} or significant interactions between 
chain segments and the cosolvent molecules \cite{rb,rkr}, its effect may 
in fact be inverted. However, because of its ubiquity as a solubilising 
agent for most binary systems, we continue to consider urea as a generic 
example of a strong chaotrope. 

The aggregation singularity was reproduced in the qualitative analysis of 
urea as a chaotropic agent performed in Ref.~\cite{moelbert2}, and is 
illustrated in Fig.~\ref{FigBoulemu_T_MF_chaot}. Because of the discrete 
nature of the Monte Carlo simulations, the exact parameters for the critical 
point cannot be found using this technique; these were instead obtained only 
within the mean-field approximation, and as a result cannot be considered 
to be quantitatively accurate for the parameters of the system under 
consideration. Figure~\ref{FigBoulemu_T_MF_chaot}(a) shows the shrinking 
of the coexistence curves with increasing cosolvent concentration until 
they vanish at a critical value $\rho_c^{\ast} = 0.81$ 
({\it cf.}~$\rho_c^{\ast} \approx 0.55$ expected from inspection of 
the Monte Carlo results in Fig.~\ref{FigBouleMC_chaot}). The 
corresponding evolution of the transition line in the $\mu$-$T$ phase 
diagram is shown in Fig.~\ref{FigBoulemu_T_MF_chaot}(b).

With the canonical choice of degeneracy parameters 
(Sec.~\ref{secModel}B) $q_{ds} = 49,\ q_{db} = 40,\ q_{ob} = 
10$ and $q_{os} = 1$~\cite{silverstein2}, the values $\eta_b = 
9.0$ and $\eta_s = 0.1$ required in eq.~(\ref{eqDegen2}) to reproduce 
this critical vanishing are indeed such that $q_{ob,c} = 1$ 
and $q_{os,c} = 0.9$ become very close. The level of ``fine-tuning'' 
of the underlying parameters required in the adapted MLG model reflects 
not a weakness of the framework but rather of the strongly chaotropic 
properties of urea in water: the destruction of hydrogen-bonded networks 
is almost complete at high concentrations. In practice, concentrated 
urea is used to destabilise aqueous solutions of many folded proteins, 
micelles and aggregates of hydrophobic particles. While we are unaware 
of systematic experimental studies of the effects of urea on the extent 
of the coexistence regime, aggregate destabilisation has been achieved 
at high concentrations in certain systems. At room temperature, the 
solubility of the highly hydrophobic amino-acid phenylalanin is doubled 
in an $8$M urea solution (a solution with equal volume fractions 
of urea and water) \cite{nozaki1}.

\subsection{Sodium Chloride as a Weak Kosmotrope}

Kosmotropic cosolvents increase the extent of water structure formation, 
raising the number of ordered bulk and shell sites. While this process 
may be more efficient for ordered shell states, their degeneracy cannot 
exceed that of ordered bulk states. A maximally kosmotropic agent is 
one which causes sufficient structural enhancement that the number 
of ordered bulk states rises to the point where it is equal to the number 
of disordered bulk states, $q_{ob} = q_{db}$. Singular behaviour for 
a strongly kosmotropic cosolvent (the counterpart to the vanishing of 
aggregation caused by chaotropic cosolvents) is then the perfect 
aggregation of the system with densities $\rho_{1} = 0$ and $\rho_{2} 
= 1$ (Fig.~\ref{FigBouleMFKosmo}), {\it i.e.}~extension of the coexistence 
regime across the entire phase diagram. Only at the lowest temperatures, 
where entropy effects are irrelevant, would the mechanism of cage 
formation allow solution of the solute particles. For a sufficiently 
strong kosmotropic effect (high concentrations of a strong agent) there 
is no reason why $q_{ob}$ should not exceed $q_{db}$, a situation which 
represents the breakdown of the sequence of energies and degeneracies 
characteristic of a system with a closed-loop miscibility curve 
(Sec.~\ref{secModel}A). Complete aggregation is the hallmark of this 
breakdown.

The ``symmetrical'' parameters $\eta_b = 9.0$ and $\eta_s = 0.1$ used 
in the analysis of Sec.~\ref{secKosmo} give the values $q_{ob} = 19$ 
and $q_{db} = 31$ [eq.~(\ref{eqDegen})]. These are clearly not sufficiently 
close to represent a strongly kosmotropic cosolvent, which would explain 
both the rather modest expansion of the coexistence regime even with 
relatively high cosolvent concentrations ($\rho_k$) and the small values 
of the upper particle density $\rho_2 < 1$ (Figs.~\ref{FigBouleMFKosmo}, 
\ref{FigBouleMCKosmo}). One may thus conclude that quantitative symmetry 
between kosmotropic and chaotropic effects cannot be expected for any 
temperatures or densities, except in the event of very carefully chosen 
degeneracy parameters. 

\begin{figure}[t!]
\includegraphics[width=8.5cm]{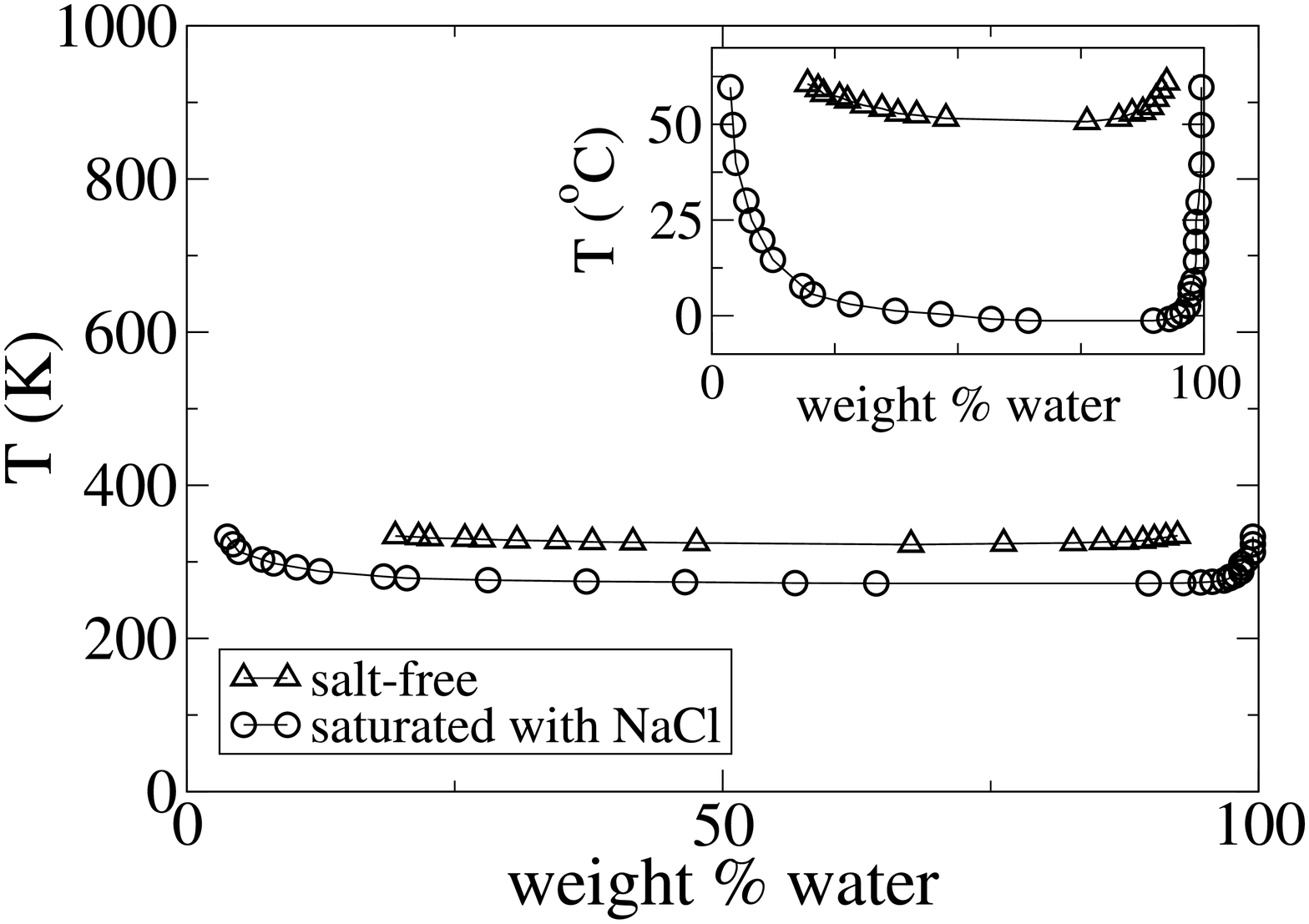}
\caption{Coexistence curve measured in the experimentally accessible 
temperature range for N,N-Diethylmethylamine in water in the presence 
and in the absence of the kosmotropic cosolvent sodium chloride, 
reproduced from Ref.~\cite{ting}. The LCST is reduced from $51^{\circ}$C 
in the salt-free case to $-0.6^{\circ}$C in the saturated cosolvent 
solution.}
\label{FigTingKosmo}
\end{figure}

To our knowledge, rather little data is presently available concerning 
the phase diagram of hydrophobic solute particles in cosolvent solutions. 
However, Ting {\it et al.}~\cite{ting} have measured the lower coexistence 
curves for N,N-Diethylmethylamine both in pure water and in a mixture 
of water and the cosolvent sodium chloride (Fig.~\ref{FigTingKosmo}), 
which is usually considered to be weakly kosmotropic. A decrease of the 
LCST from $51^{\circ}$C to $-0.6^{\circ}$C is observed for a solution 
saturated with sodium chloride, corresponding to extension of the 
aggregation regime over a significantly greater temperature range. 
Although the data do not extend to sufficiently high temperatures to 
discuss the full extension of the aggregation regime in density 
(Fig.~\ref{FigTingKosmo}, inset), the qualitative similarity of 
Figs.~\ref{FigBouleMFKosmo} and \ref{FigBouleMCKosmo} to the 
experimentally determined coexistence curves (Fig.~\ref{FigTingKosmo}, 
main panel) suggests that sodium chloride would be appropriately 
classified as a ``weak'' kosmotrope in its effect on 
N,N-Diethylmethylamine. While its activity is significantly 
stronger (or its saturation concentration significantly higher) than 
that of sodium sulphate, investigated by the same authors \cite{ting}, 
at maximal concentrations it appears that the aggregation singularity 
is not reached and the closed-loop form of the miscibility curve is 
preserved. Here we note that, while the kosmotropic action of sodium 
chloride is due to its ionic nature in solution, the considerations of 
the adapted MLG framework (Sec.~\ref{secModel}B) remain valid to model 
this effect. A more detailed comparison of the model results with the 
data is precluded by the fact that the cosolvent concentration in the 
saturated solution changes with both temperature and solute density. 
Measurements of different solutions are required to characterise the 
dependence of aggregation enhancement on the molecular properties of 
the solute and on both solute and cosolvent concentrations in the system. 

Finally, preferential exclusion of kosmotropic cosolvents from the 
immediate vicinity of hydrophobic particles in aqueous solutions is 
measured only for very high cosolvent concentrations. Concentrations 
of compatible osmolytes, such as sucrose and betaine in the cytoplasm, 
typically reach values well in excess of $0.5$M. The stability of the 
protein lactate dehydrogenase against thermal denaturation increases 
by approximately $90\%$ in a $1$M sucrose solution at room temperature,
and by $100\%$ for the same concentration of hydroxyecotine~\cite{galinski2}, 
classifying these cosolvents as being strongly kosmotropic. In the current 
model it is difficult to specify the exact cosolvent concentration because 
one site contains a group of water molecules, but the results are 
nevertheless qualitatively consistent with observation. We have found 
that, except for specifically chosen values of the parameters $q_\sigma$ 
and $\eta_\sigma$, cosolvent concentrations well in excess of $10\%$ are 
required to show a clear stabilising effect on hydrophobic aggregates 
(Fig.~\ref{FigBouleMCKosmo}).

\section{Summary}
\label{secConclusion}

We have discussed the physical mechanism underlying the action of 
kosmotropic and chaotropic cosolvents on hydrophobic aggregates in 
aqueous solutions. The aggregation phenomena arise from a balance 
between enthalpic and entropic contributions to the solvent free 
energy, which is fully contained within an adapted version of the 
bimodal model of Muller, Lee and Graziano. This model considers two 
populations of water molecules with differing physical properties, 
and generates indirectly the effective hydrophobic interaction 
between solute particles. Such a description, based on the microscopic 
details of water structure formation around a solute particle, is 
expected to be valid for small solute species, characterised by 
lengthscales below 1 nm, and for those larger solute particles whose 
solubility is dominated by the atomically rough (highly curved) nature 
of their surfaces. 

In the adapted MLG framework, one may include the ability of kosmotropic 
cosolvents to enhance and of chaotropic cosolvents to disrupt the 
structure of liquid water simply by altering the state degeneracies 
[eqs.~(\ref{eqDegen}) and (\ref{eqDegen2})] associated with the two 
populations characterising this structure. As a consequence only of 
these changes in the properties of the solvent, the model reproduces 
all of the physical consequences of cosolvents in solution: for 
kosmotropes the stabilisation of aggregates, expansion of the 
coexistence regime and preferential cosolvent exclusion from the 
hydration shell of hydrophobic solute particles; for chaotropes the 
destabilisation of aggregates, a contraction of the coexistence regime 
and preferential binding to solute particles. 

The microscopic origin of these preferential effects lies in the 
energetically favourable enhancement of hydrogen-bonded water structure 
in the presence of a kosmotropic cosolvent, and conversely the avoidance 
of its disruption in the presence of a chaotrope. In a more macroscopic 
interpretation, the preferential hydration of solute particles by 
kosmotropic agents can be considered to strengthen the solvent-induced, 
effective hydrophobic interaction between solute particles, thus
stabilising their aggregates; similarly, the preferential binding 
of chaotropic agents to solute particles reduces the hydrophobic 
interaction, causing the particles to remain in solution. The essential 
cosolvent phenomena may thus be explained purely by the propensity of 
these agents to promote or suppress water structure formation, and the 
effectiveness of a cosolvent is contained in a transparent way in the 
microscopic degeneracy parameters of the solvent states.  

Our focus on water structure is intended to extract one of the primary 
factors determining aggregation and cosolvent activity, and is not 
meant to imply that other interactions are not important. The range of 
solute species and additional contributions which are beyond the scope 
of our analysis includes extremely small solutes, large, atomically 
flat solute surfaces, solutes with differing surface hydrophobicity 
(such as polar side-groups) and surfaces undergoing specific chemical 
(or other) interactions with the cosolvent. However, from the 
qualitative trends, and even certain quantitative details, observed 
in a wide variety of aqueous systems it appears that the adapted MLG 
model is capable of capturing the essential phenomenology of cosolvent 
activity on aggregation and solubility for solutes as diverse as short 
hydrocarbons and large proteins. 

\acknowledgments

We thank the Swiss National Science Foundation for financial support 
through grants FNRS 21-61397.00 and 2000-67886.02.


\begin{thebibliography}{99}

\bibitem{galinski2}
E.A. Galinski, Compatible solutes of halophilic eubacteria: molecular 
principles, water-solute interaction, stress protection, Experientia 
{\bf 49} (1993) 487-496.

\bibitem{yancey}
P.H. Yancey, M.E. Clark, S.C. Hand, R.D. Bowlus and G.N. Somero, 
Living with water stress: evolution of osmolyte systems, Science 
{\bf 217} (1982) 1214-1222.

\bibitem{somero}
G.N. Somero, Protons, osmolytes, and fitness of internal milieu for 
protein function, Am. J. Physiol. {\bf 251} (1986) R197-R213.

\bibitem{wiggins}
P.M. Wiggins,  Role of water in some biological processes, Microbiol. 
Rev. {\bf 54} (1990) 432-449.

\bibitem{galinski}
E. A. Galinski, M. Stein, B. Amendt and M. Kinder, The kosmotropic 
(structure-forming) effect of compensatory solutes, Comp. Biochem. 
Physiol. {\bf 117A} (1997) 357-365.

\bibitem{arakawa}
T. Arakawa and S.N. Timasheff, Mechanism of poly(ethylene glycol) 
interaction with proteins, Biochemistry {\bf 24} (1985) 6756-6762.

\bibitem{timasheff3}
S.N. Timasheff, in: Water and life, ed. G.S. Somero (Springer Verlag, 
Berlin, 1992) p.~70.

\bibitem{rszjk} T. S\"oderlund, K. Zhu, A. Jutila and P.K.J. Kinnunen, 
Effects of betaine on the structural dynamics of {\em Thermomyces 
(Humicola) lanuginosa} lipase, Coll. Surf. B: Biointerfaces {\bf 26} 
(2003) 75-83.

\bibitem{rbb}
D.W. Bolen and I.V. Baskakov, The osmophobic effect: natural selection 
of a thermodynamic force in protein folding, J. Mol. Biol. {\bf 310} 
(2001), 955-963.

\bibitem{schellman}
J.A. Schellman, Selective binding and solvent denaturation, Biopolymers 
{\bf 26} (1987) 549-559; A simple model for solvation in mixed solvents, 
Biophys. Chem. {\bf 37} (1990) 121-140; The relation between the free 
energy of interaction and binding, Biophys. Chem. {\bf 45} (1993) 273-279.

\bibitem{timasheff2}
S.N. Timasheff, Thermodynamic binding and site occupancy in the light 
of the Schellman exchange concept, Biophys. Chem. {\bf 101-102} (2002) 
99-111.

\bibitem{tovchi}
A. Tovchigrechko, M. Rodnikova and J. Barthel, Comparative study of 
urea and tetramethylurea in water by molecular dynamics simulations, 
J. Mol. Liq. {\bf 79} (1999) 187-201.

\bibitem{oro}
J.R. De Xammar Oro, Role of co-solute in biomolecular stability: glucose, 
urea and the water structure, J. Biol. Phys. {\bf 27} (2001) 73-79.

\bibitem{feng}
Y. Feng, Z.-W. Yu and P.J. Quinn, Effect of urea, dimethylurea, and 
tetramethylurea on the phase behavior of dioleoylphosphatidylethanolamine, 
Chem. Phys. Lipids {\bf 114} (2002) 149-157.

\bibitem{plumridge}
T.H. Plumridge and R.D. Waigh, Water structure theory and some implications 
for drug design, J. Pharm. Pharmacol. {\bf 54} (2002) 1155-1179.

\bibitem{nozaki1}
Y. Nozaki and  C. Tanford, The solubility of amino acids and related 
compounds in aqueous urea solutions, J. Biol. Chem. {\bf 238} (1963) 
4074-4081.

\bibitem{nozaki1a}
Y. Nozaki and  C. Tanford, The solubility of amino acids, diglycine, and 
triglycine in aqueous guanidine hydrochloride solutions, J. Biol. Chem. 
{\bf 245} (1970) 1648-1652.

\bibitem{nandi}
P.K. Nandi and  D.R. Robinson, Effects of urea and guanidine hydrochloride 
on peptide and non-polar groups, Biochemistry {\bf 23} (1984) 6661-6668.

\bibitem{walrafen}
G.E. Walrafen, Raman spectral studies of the effects of urea and sucrose 
on water structure, J. Chem. Phys. {\bf 44} (1966) 3726-3727.

\bibitem{wetlaufer}
D.B. Wetlaufer, S.K. Malik, L. Stoller and R.L. Coffin, Nonpolar group 
participation in the denaturation of proteins by urea and guanidinium 
salts, J. Am. Chem. Soc. {\bf 86} (1967) 508-514.

\bibitem{roseman}
M. Roseman and W.P. Jencks, Interactions of urea and other polar 
compounds in water, J. Am Chem. Soc. {\bf 97} (1975) 631-640.

\bibitem{kita}
Y. Kita, T. Arakawa, T.-Y. Lin and S.N. Timasheff, Contribution of the 
surface free energy perturbation to protein-solvent interactions, 
Biochemistry {\bf 33} (1994) 15178-15189.

\bibitem{franks}
F. Franks, Protein stability: the value of 'old literature', Biophys. 
Chem. {\bf 96} (2002) 117-127, and references therein.

\bibitem{creighton}
S.N. Timasheff and T. Arakawa, in: Protein structure: a practical 
approach, ed. T.E. Creighton (IRL Press, Oxford, 1997) p.~331.

\bibitem{frank} 
H.S. Frank and M.W. Evans, Free volume and entropy in condensed systems, 
J. Chem. Phys. {\bf 13} (1945) 507-532.

\bibitem{yaacobi}
M. Yaacobi and A. Ben-Naim, Solvophobic interaction, J. Phys. Chem. 
{\bf 78} (1974) 175-178.

\bibitem{privalov}
P.L. Privalov and S.J. Gill, Stability of protein structure and 
hydrophobic interaction, Adv. Protein Chem. {\bf 39} (1988) 191-234.

\bibitem{deJong}
P.H.K. De Jong, J.E. Wilson and G.W. Neilson,  Hydrophobic hydration of 
methane, Mol. Phys. {\bf 91} (1997) 99-103.

\bibitem{pertsemil}
A. Perstemilidis, A.M. Saxena, A.K. Soper, T. Head-Gordon and R.M. Glaeser, 
Direct evidence for modified solvent structure within the hydration shell 
of a hydrophobic amino acid, Proc. Natl. Acad. Sci. USA {\bf 93} (1996) 
10769-10774.

\bibitem{kauzmann}
W. Kauzmann, Some factors in the interpretation of protein denaturation, 
Adv. Protein Chem. {\bf 14} (1959) 1-63.

\bibitem{tanford} 
C. Tanford, The hydrophobic effect: formation of micelles and biological 
membranes (Wiley, New York, 1980).

\bibitem{stillinger80}
F.H. Stillinger, Water revisited, Science {\bf 209} (1980) 451-457.

\bibitem{ben-naim2} 
A. Ben-Naim, Water and aqueous solutions: introduction to a molecular 
theory (Plenum Press, New York, 1974). 

\bibitem{ludwig}
R. Ludwig, Water: from clusters to the bulk, Angew. Chem. Int. Ed. Engl. 
{\bf 40} (2001) 1808-1827.

\bibitem{timasheff}
S.N. Timasheff, Protein-solvent preferential interactions, protein 
hydration, and the modulation of biochemical reactions by solvent 
components, Proc. Natl. Acad. Sci. {\bf 99} (2002) 9721-9726. 

\bibitem{moelbert2}
S. Moelbert and P. De Los Rios, Chaotropic effect and preferential binding 
in a hydrophobic interaction model, J. Chem. Phys. {\bf 119} (2003) 7988-8001.

\bibitem{muller} 
N. Muller, Search for a realistic view of hydrophobic effects, Acc. 
Chem. Res. {\bf 23} (1990) 23-28.

\bibitem{lee}
B. Lee and G. Graziano, A two-state model of hydrophobic hydration that 
produces compensating enthalpy and entropy changes, J. Am. Chem. Soc. 
{\bf 22} (1996) 5163-5168.

\bibitem{silverstein1} 
K.A.T. Silverstein, A.D.J. Haymet and K.A. Dill, The strength of hydrogen 
bonds in liquid water and around non-polar solute, J. Am. Chem. Soc. 
{\bf 122} (2000) 8037-8041.

\bibitem{moelbert1}
S. Moelbert and P. De Los Rios, Hydrophobic interaction model for upper 
and lower critical solution temperatures, Macromolecules {\bf 36} (2003) 
5845-5853.

\bibitem{silverstein2} 
K.A.T. Silverstein, A.D.J. Haymet and K.A. Dill, Molecular model of 
hydrophobic solvation, J. Chem. Phys. {\bf 111} (1999) 8000-8009.

\bibitem{paschek} D. Paschek, Temperature dependence of the hydrophobic 
hydration and interaction of simple solutes: an examination of five popular 
water models, J. Chem. Phys. {\bf 120} (2004) 6674-6690; Heat capacity 
effects associated with the hydrophobic hydration and interaction of simple 
solutes: a detailed structural and energetical analysis based on molecular 
dynamics simulations, J. Chem. Phys. {\bf 120} (2004) 10605-10617.

\bibitem{delos2}
P. De Los Rios and G. Caldarelli, Cold and warm swelling of hydrophobic 
polymers, Phys. Rev. E {\bf 63} (2001) 031802-1-5.

\bibitem{delos} 
P. De Los Rios and G. Caldarelli, Putting proteins back into water, 
Phys. Rev. E {\bf 62} (2000) 8449-8452.

\bibitem{caldarelli}
G. Caldarelli and P. De Los Rios, Cold and warm denaturation of proteins, 
J. Biol. Phys. {\bf 27} (2001) 229-241.

\bibitem{rmnd}
S. Moelbert, B. Normand and P. De Los Rios, Solvent-induced micelle 
formation in a hydrophobic interaction model, Phys. Rev. E {\bf 69} 
(2004) 061924-1-11.

\bibitem{rlmr}
C.Y. Lee, J.A. McCammon and P.J. Rossky, The structure of liquid water 
at an extended hydrophobic surface, J. Chem. Phys. {\bf 80} (1984)
4448-4453.

\bibitem{rhea}
X. Huang, C.J. Margulis and B.J. Berne, Do molecules as small as neopentane 
induce a hydrophobic response similar to that of large hydrophobic surfaces ? 
J. Phys. Chem. B {\bf 107} (2003) 11742-11748.

\bibitem{rsc}
S. Shimizu and H.S. Chan, Origins of protein denatured states compactness 
and hydrophobic clustering in aqueous urea: inferences from nonpolar 
potentials of mean force, Proteins: Struct. Funct. Genet. {\bf 49} 
(2002) 560-566.

\bibitem{rlcw}
K. Lum, D. Chandler and J.D. Weeks, Hydrophobicity at small and large 
length scales, J. Phys. Chem. {\bf 103} (1999), 4570-4577.

\bibitem{finer}
E.G. Finer, F. Franks and M.J. Tait, Nuclear magnetic resonance studies 
of aqueous urea solutions, J. Am. Chem. Soc. {\bf 94} (1972) 4424-4427.

\bibitem{rb}
J.M. Scholtz, D. Barrick, E.J. York, J.M. Stewart and R.L. Baldwin, Urea 
unfolding of peptide helices as a model for interpreting protein unfolding, 
Proc. Natl. Acad. Sci. USA {\bf 92} (1995) 185-189.

\bibitem{rkr}
R.A. Kuharski and P.J. Rossky, Solvation of hydrophobic species in aqueous 
urea solution: a molecular dynamics study, J. Am. Chem. Soc. {\bf 106} 
(1984) 5794-5800.

\bibitem{rg}
G. Graziano, On the solubility of aliphatic hydrocarbons in 7 M aqueous 
urea, J. Phys. Chem. B {\bf 105} (2001) 2632-2637.

\bibitem{ting}
A.M. Ting, S. Lynn and J.M. Prausnitz, Liquid-liquid equilibria for aqueous 
systems containing N,N-Diethylmethylamine and sodium chloride or sodium 
sulfate, J. Chem. Eng. Data {\bf 37} (1992) 252-259.

\end{thebibliography}
\end{document}